# Observation of vacancy-related polaron states at the surface of anatase and rutile TiO$_2$ by high-resolution photoelectron spectroscopy

*Mark J. Jackman[a], Peter Deák[b], Karen L. Syres[c], Johan Adell[d], Balasubramanian Thiagarajan[d], Anna Levy[e] and Andrew G. Thomas[a,\*]*

[a] School of Physics and Astronomy and Photon Science Institute, Alan Turing Building, The University of Manchester, Oxford Road, Manchester, M13 9PL, UK

[b] Bremen Center for Computational Materials Science, University of Bremen, PoB 330440, D-28334 Bremen, Germany

[c] School of Chemistry, The University of Nottingham, University Park, Nottingham, NG7 2RD, UK

[d] MaxLab, Ole Römers Väg, 223 63 Lund, Sweden

[e] Synchrotron Soleil, L'Orme des Merisiers, Saint Aubin BP 48 – 91192- Gif-sur-Yvette Cedex, France[+]

\*corresponding author: a.g.thomas@manchester.ac.uk

[+]Current address: Sorbonne Universités, UPMC, Paris 06, CNRS, INSP, UMR 7588, INSP, F-75005, Paris, France

**Abstract:** Defects in the surface region of a reducible oxide, as TiO$_2$, have a profound effect on applications, while their nature is very much influenced by the possibility of small polaron formation. Here, we probe rutile (110) and anatase (101) single crystals via high-resolution ultraviolet photoelectron spectroscopy and resolve multiple components of the well-known defect state in the band gap. In rutile, we find two at VBM+2.1 eV and VBM+1.4 eV, which we assign to *subsurface* polaron traps and vacancy-bound states, respectively, confirming the predicted partial suppression of polaron formation at high vacancy concentration. New defects are created *in situ* on the anatase surface by the synchrotron beam. We assign a component at VBM+2.3 eV, which can be removed by annealing, to polaron states associated with *surface* oxygen vacancies. We also identify a second component at VBM+1.6 eV, which can not be removed by annealing, and is too deep to be associated with oxygen vacancies.

In highly polar crystals with strong electron-phonon coupling (such as many II-VI semiconductors, alkali halides, oxides, etc.), a charge carrier is dressed in a cloud of virtual phonons, and that can be considered together with the carrier as a new composite particle, called a polaron. If the phonon cloud corresponds to a strong local relaxation of the lattice, the carrier is essentially trapped in a so-called "small polaron" state, and transport can proceed only through thermally activated hopping. Such carrier-trapping is obviously a critical issue in many application of these materials. Theoretical description of small polaron states has become possible only recently, with corrections for the electron self-interaction error of standard density functional theory (by using Hubbard-$U$ terms [1] or hybrid functionals [2]), but comparison with the relatively scarce experimental data is still controversial. $TiO_2$ is an excellent model material for investigating the possibility and consequences of small polaron states. $TiO_2$ is used for photo-catalysis [3] or in electrochemical solar cells [4], and it is being considered as transparent conducting oxide (TCO) [5] or phase change material for resistive memory (RRAM) [6] applications. It is also a prototype reducible oxide, and the interaction between carrier self-trapping by small polaron states and self-doping, e.g., by oxygen vacancies ($V_O$), gives rise to an exceptionally rich physics in this material. Traditionally, the (110) surface of rutile (R) has been most often studied, even though anatase (A), having (101) as the most stable surface, is catalytically more active [7].

According to theoretical predictions [8,9,10], the rutile polymorph of $TiO_2$ has the capacity for electron self-trapping in small polaron states. This prevents effective *n*-type doping (and the use as TCO) and limits mobility, in contrast to the anatase polymorph with no small electron-polaron states in the bulk (and excellent electron transport properties in electrochemical solar cells). This difference has been confirmed experimentally [11]. Very recently, however, theoretical work has indicated that small electron-polaron states can exist on the anatase [101] surface as well [12]. It has long been postulated that defects, like Vo, at the surface are catalytic centres in $TiO_2$. The presence of polaronic traps greatly influences not only the electronic and optical properties of $V_O$ [9] but also its preferred position. In principle, $V_O$ is a double donor, because ejection of the neutral oxygen leaves two electrons. These can stay in the vacancy (in deep, effective-mass-like states) or convert nearby $Ti^{4+}$ atoms into $Ti^{3+}$, accompanied by strong relaxation of the neighbourhood, i.e., forming a small polaron. It has been predicted theoretically [13,14,15] and confirmed experimentally [16,17] that in [110] oriented rutile $V_O$ prefers surface (so-called "bridging") sites but automatically loses both of its electrons to subsurface polaron traps. Theoretical work has indicated, however, that increasing the vacancy concentration, the possibility of polaron formation can be suppressed, and $V_O$ in rutile may retain one or two electrons [9]. In contrast to rutile, experiments in anatase indicate that $V_O$ prefers subsurface sites [18,19], where they retain their electrons in deep effective-mass-like orbitals. However, a most recent theoretical work came to the conclusion, that the preference for subsurface sites can only be explained if the Fermi-level is pinned below the ionization energy, while neutral $V_O$ is thermodynamically more stable on the surface [12]. Obviously, more work is needed here, since the question of surface vs. subsurface vacancies has profound consequences for catalysis, because subsurface vacancies can be active carrier traps while those on the surface can be passivated by adsorbates [18].

Whether electrons are in donor-like vacancy or in small polaron trap states, whether localized on or below the surface, they can be excited into the conduction band by a characteristic energy. Ultraviolet photoelectron spectroscopy (UV-PES), probing both the surface and the immediate subsurface region could, in principle, provide information about the existence and site-preference of the various vacancy-related and polaron states in the two polymorphs of $TiO_2$, and contribute to the understanding of polaron formation. Surprisingly, however, the spectra usually reveal only one defect-related state with an energy of about 1 eV from the conduction band edge, irrespective of the polymorph and of the surface orientation [7,20,21,22]. This state is mostly attributed to electrons in small polaron states in conjunction with subsurface Ti interstitials [23] or surface oxygen vacancies [16]. It has been argued that, except for very strongly reduced samples, $V_O$ dominates [24]. In a recent study, however, a second state has been observed in R-$TiO_2$ (110) [25]. Considering the variety of possibilities listed in the previous paragraph, it could be informative to look at the defect band in high resolution.

To shed light on the details of small polaron formation, we probe the bandgap state observed in the valence band photoemission spectrum, via synchrotron radiation, of both R-$TiO_2$ (110) and A-$TiO_2$ (101), and take advantage of a high-resolution analyzer. Comparison with theory shows that the main component has polaronic origin in both cases, at positions which are in excellent agreement with theoretical predictions on near-surface polarons. In rutile we also resolve a component that can be related to electrons *inside* $V_O$. This finding corroborates the theoretical prediction that high $V_O$ concentration partially suppresses small polaron formation. In anatase no subsurface $V_O$ component could be resolved, even after sputtering and annealing.

This work was carried out on the undulator beamline I4 ($13 \leq h\nu \leq 200$ eV) at MAX-lab, Sweden. The I4 endstation is equipped with a PHOIBOS 100 mm CCD analyzer (SPECS). Measurements on the A-$TiO_2$(101) surface were repeated on the Antares beamline at Synchrotron Soleil ($12 \leq h\nu \leq 1000$ eV), equipped with a Scienta R4000 hemispherical analyser. The A-$TiO_2$ (101) sample (5 mm$^2$, Pi-kem Ltd.) was cut from a natural single crystal. The R-$TiO_2$ (110) sample was a grown crystal (10 mm$^2$, Pi-kem Ltd). The crystals were held in place on Mo sample plates by two strips of tantalum wire. The crystal was cleaned by repeated 1 keV Ar$^+$ ion bombardment and annealing to 700 °C (measured by a pyrometer) in vacuum, until sharp (1 x 1) LEED patterns were obtained [26,27]. Bandgap state photoemission spectra were recorded at normal emission at 50 eV photon energy and aligned relative to the valence band maximum (VBM). A Tourgaard background was subtracted from the bandgap state spectra as it gave the best fit. Voigt curves (70% Gaussian: 30% Lorentzian) were used to fit the spectra [28]. The sample holder and manipulator were cooled with liquid nitrogen to 100-120 K.

Figure 1 shows the bandgap state of R-$TiO_2$ (110). The complete valence band spectrum is shown in the Supplementary Information S.1. (see below), and is in good agreement with previous data [22]. The photon energy is at 50 eV, close to the Ti 3p–3d resonance of titanium (47 eV) [26].

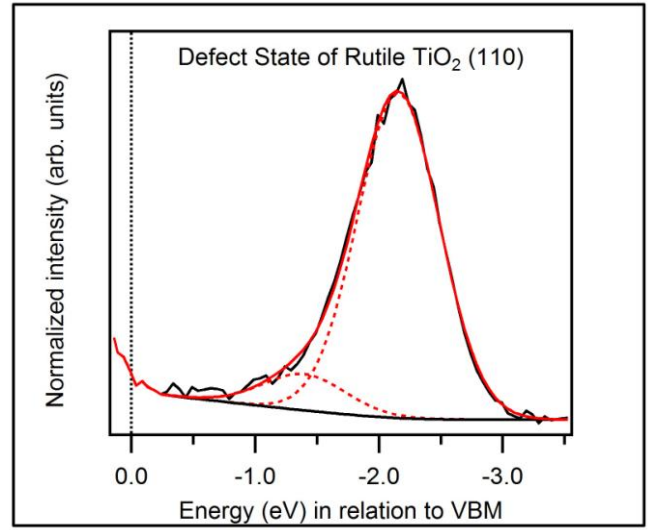

*Figure 1: The defect peak of Rutile TiO$_2$ (110) recorded at 50 eV photon energy.*

The energies of the peaks with respect to the VBM are 1.4 eV (binding energy: 1.7 eV) and 2.1 eV (binding energy: 1.0 eV) in a ratio of 10:90. The spectrum does not change over several hours under the beam (see below for Supplementary Information S.2.). The defect density at the surface was estimated at ~7 % from the Ti$^{3+}$ contributions to the Ti 3p spectrum (see below for Supplementary Information S.3.). As to the origin of the two peaks in the photoelectron spectrum, these states are not likely to arise through impurities in the lattice as no impurities were detected in the photoemission spectra or when the crystal was analysed by X-ray fluorescence (XRF) (limit ~0.1 %). Theoretical studies using the HSE06 hybrid functional [9,13] suggest that V$_O$, both in the bulk and on the surface of rutile, automatically loses both of its electrons in favour of Ti(4+/3+) traps with bulk-like environment. The latter give rise to band gap states about 2.1-2.2 eV above the VBM. However, it has also been shown [9] that, as the concentration of such filled trap states increases, V$_O$-s in the bulk (or subsurface) may retain their electrons, giving rise to band gap states at +2.5 eV from VBM, attributed to V(0/+), and +1.5 eV from VBM, attributed to the second ionisation potential, V(+/2+). Considering that the thermal energy required for the first ionization of such a vacancy state was predicted to be low [9,10], and the fact that the spectrum does not change under the beam, the two components in our measurement can be interpreted in terms of electrons in subsurface polaron states (first peak at +2.1 eV from VBM) and in subsurface, singly-ionized vacancy-bound states (peak at VBM+1.4 eV), in close agreement with the predicted theoretical values. (N.B.: subsurface vacancies have been already observed in R(110) samples [29].) The much smaller size of the peak at VBM+1.4 eV can then be explained, because the second electron would need to be ejected before any recombination occurs, so this state has a smaller population. The detection of this second ionization energy is a strong indication for the concentration dependence of the vacancy behaviour [9]. Multiple states in the bandgap region have been recently observed by Sánchez-Sánchez *et al.* [25] at 1.17 eV and 0.75 eV binding energy. They attribute the smaller peak at 1.17 eV to a state induced by a (1 x 2) reconstruction on the surface. It is known that this reconstruction occurs in heavily bulk-reduced R-TiO$_2$ (110) single crystals [7], thus the origin of the second defect-related peak observed in such crystals [25] may as well lie in the high Vo concentration and arise from the second ionisation potential as well. It is also possible, however, that at higher defect concentrations Ti interstitials in the subsurface region contribute to the peak at VBM+1.4 eV.

Figure 2 shows the photoelectron spectra of the bandgap state of A-TiO$_2$ (101). The valence band spectrum is included in the Supplementary Information S.1 (see below), and is in good agreement with previous data [22]. The spectrum recorded for the bandgap state of A-TiO$_2$ (101) is clearly asymmetric and can only be fitted with a minimum of three peaks. As can be seen in Figure 2, the structure and intensity of the defect peak changes under the beam. Therefore, to determine the energy of the peaks, an average of sixteen spectra were taken (data table in Supplementary Data S.4 below) to give energies, with respect to the VBM, of +2.3 eV (binding energy: 1.0 eV), +1.6 eV (binding energy: 1.7 eV) and +0.8 eV (binding energy: 2.4 eV).

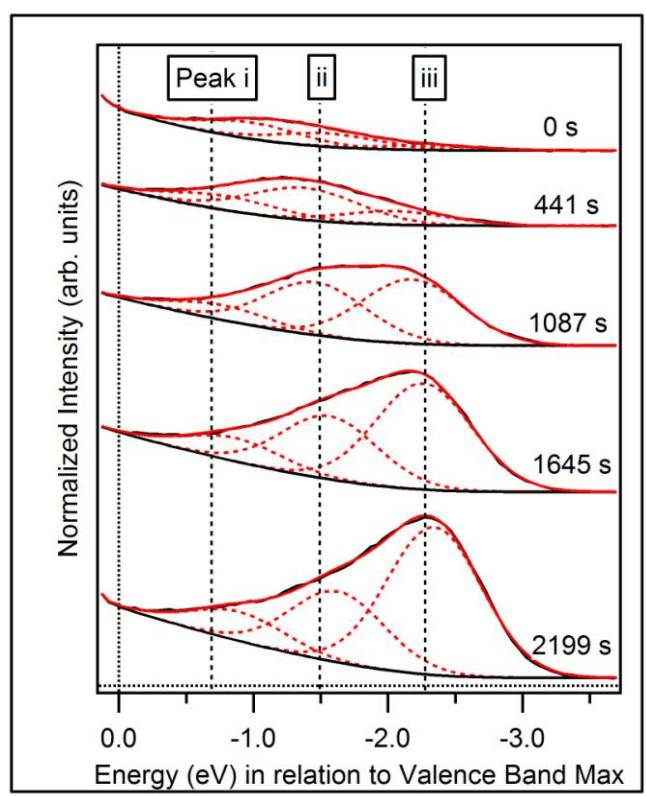

*Figure 2: The defect peak of A-TiO$_2$ (101) recorded at 50 eV photon energy shows the change in the defect-peak structure over time See text for information on energy level*

For the A-TiO$_2$ (101) surface, it was not possible to determine the ratio of the features contributing to the defect state since they were changing under the beam, and no equilibrium was reached on the timescale of the measurement. The beam-induced changes, however, give an insight into the nature of the resolved components. The intensity of the peak at VBM+0.8 eV remains relatively constant [30], and we believe this to be due to an impurity. Ta and Nb impurities were determined by XRF to both be present at ~3000 ppm, although neither Ta nor Nb were observed in X-ray photoelectron spectroscopy (XPS) data. The Ta is thought to arise through contact with the Ta sample plate during annealing. Nb is an impurity in the natural crystal. However, these metals are shallow electron donors in anatase, with levels near the conduction band edge [5,8,11]. The energy position would suggest the peak at VBM+0.8 eV to be associated with a hole trap, possibly due to C impurities [31], despite the XPS spectrum indicating a surface free of contaminants.

The other two peaks fitted in Figure 2 show a change in intensity with time under the beam. The peak at VBM+1.6 initially grows rapidly and then seems to saturate. The peak at VBM+2.3, on the other hand continues to grow, and over 35 minutes has never reached saturation. An increase in intensity in this energy range under the beam has been observed just recently [32]. Figure 3a shows the effect of moving to a fresh position on the sample. The intensity is normalised to the area of the VBM+1.6 eV peak. It can be seen that the peak at VBM+2.3 eV is greatly reduced. The reduction of the two peaks indicates they are both the product, or part-product of localised beam damage on the surface. The intensity of the peak at VBM+0.8 eV remains similar, as would be expected of a subsurface impurity state. Figure 3b shows the effect of $Ar^+$ ion sputtering an $A-TiO_2$ (101) crystal. The peak shape is similar to that caused by beam damage (Figure 3a). Our data show that annealing the surface results in almost complete removal of the component at VBM+2.3 eV with very little change in the intensity of the VBM+1.6 eV peak.

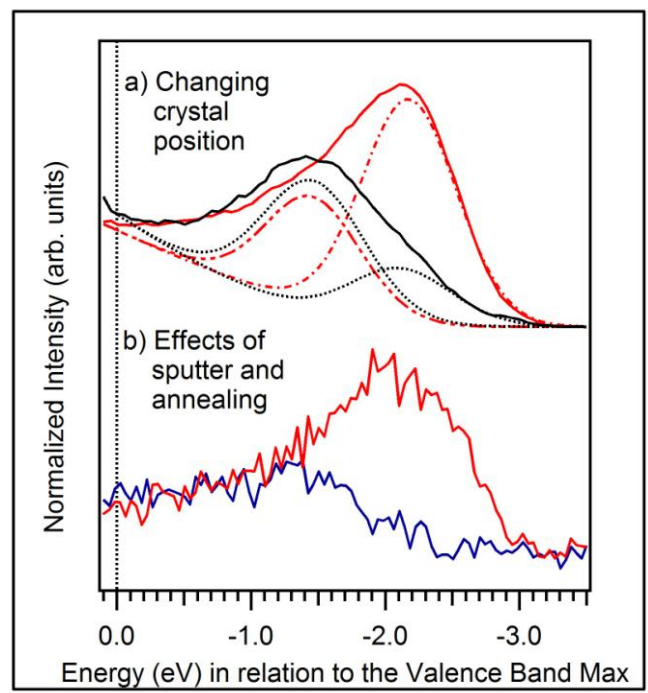

*Figure 3: Surface vs. subsurface defects on $A-TiO_2$ (101). a) Shows the effects of moving the beam. The red spectrum is after a period of time under the beam, and the black spectrum, (normalised to the peak at VBM+1.6 eV) was taken directly after moving the crystal (VBM+0.8 eV peak omitted for clarity). b) Band-gap spectra taken on the bending magnet beamline D1011 on MAX II, MAX-lab at 47 eV photon energy. The red line is the spectrum*

*after sputtering and the blue line shows the removal of these defects after annealing. This bending magnet beamline did not create defects.*

The data in Figure 3b may suggest at first sight that the peak at VBM+1.6 eV could be associated with subsurface defects. However, the changes in intensity of the VBM+1.6 eV peak under the beam does not appear consistent with subsurface behaviour since one would naturally assume that the beam damage would occur more rapidly on the uppermost surface. Recent calculations by Deák *et al.* [12] suggest that such a deep state cannot be assigned to $V_O$-s either subsurface or on the surface. Very recent work by Setvin *et al.* have shown a deeper component in the defect state that they describe as a pronounced tail. Using STM, they assign this to charge trapping at step edges [32]. Adopting this interpretation would explain why this peak can not be removed by the annealing process, whereas Vo can. A change in intensity could be through localised heating by the beam driving electrons to the surface. This would also explain the saturation of the feature at VBM+1.6 eV, as physical defects are not being created, but the step edge is acting as a trap.

With regard to the component at VBM+2.3 eV, calculations by Deák *et al.* suggest that in bulk anatase the electrons are confined *inside* the $V_O$, so the band gap state is composed of two states, one lying at VBM+3.1 eV, attributed to V(0/+), and another at VBM+2.3 eV attributed to V(+/2+) [9]. Due to the very low thermal ionization energy of vacancies in anatase, (40 meV [33]), they are likely to be singly ionized, therefore the 3.1 eV peak is not observed in photoelectron spectra. The second ionization energy at VBM+2.3 eV would give rise to a weak peak (as for R-$TiO_2$ (110)). Recent calculations [12], however, show a polaronic nature for the surface vacancies, with an ionization energy of VBM+2.3 eV (by coincidence, the same as the second ionization energy of the bulk vacancy). We believe, that the peak measured at VBM+2.3 eV can be assigned to polaronic states associated with surface vacancies (polaronic states associated with subsurface vacancies were predicted to be shallower than that [12], and they would be immune to annealing).

In summary it appears, the nature of the band-gap state of $TiO_2$ is more complex than previously thought. In the case of natural A-$TiO_2$ (101), it is possible to resolve two surface-related intrinsic components. The position of one of these at VBM+2.3 eV is in excellent agreement with recent theoretical predictions for the energy of a polaron state related to a surface oxygen vacancy. The origin of the other surface related peak at VBM+1.6 eV is less clear, but assignment to step edges acting as traps [32] fits well with our results. The band gap state of R-$TiO_2$ (110) can also be deconvoluted into two components which have also been seen on a rutile $TiO_2$ (110) (1x2) reconstructed surface. Comparison with theory indicates that the component at VBM+1.4 eV can be related to electrons *inside* singly ionized subsurface vacancies, while the component measured at VBM+2.1 eV is associated with subsurface polaron states. So, despite the near-coincidence of the main components in R-(110) and A-(101), the underlying physics is qualitatively different, coming from subsurface polaron states in rutile and from surface polaron states in anatase.

**References**

[1] S. Lany and A. Zunger, Phys. Rev. B **80**, 085202 (2009).
[2] P. Deák, B. Aradi, T. Frauenheim, E.Janzén, and A. Gali, Phys. Rev. B **81**, 153203 (2010).
[3] A. Fujishima and K. Honda, Nature **238**, 37 (1972).
[4] B. Oregan and M. Gratzel, Nature **353**, 737 (1991).
[5] Y. Furubayashi, T. Hitosugi, and T. Hasegawa, Appl. Phys. Lett. **88**, 226103 (2006).
[6] D.-H. Kwon, K. M. Kim, J. H. Jang, J. M. Jeon, M. H. Lee, G. H. Kim, X.-S. Li, G.-S. Park, B. Lee, S. Han, M. Kim, and C. S. Hwang, Nature Natontech. **5**, 148 (2010).
[7] U. Diebold, Surf. Sci. Rep. **48**, 53 (2003).
[8] P. Deák, B. Aradi, and T. Frauenheim, Phys. Rev. B **83**, 155207 (2011).
[9] P. Deák, B. Aradi, and T. Frauenheim, Phys. Rev. B **86**, 195206 (2012).
[10] A. Janotti, C. Franchini, J. B. Varley, G. Kresse, and C. G. Van de Walle, Phys. Status Solidi - Rapid Res. Lett. **7**, 199 (2013).
[11] S. X. Zhang, D.C. Kundaliya, W. Yu, S. Dhar, S. Y. Young, L. G. Salamanca-Riba, S. B. Ogale, R. D. Vispute, and T. Venkatesan, J. Appl. Phys. **102**, 013701 (2007).
[12] P. Deák, B. Aradi, and T. Frauenheim, Phys. Stat. Sol. RRL **DOI 10.1002/pssr.201409139**
[13] T. Shibuya, K. Yasuoka, S. Mirbt, and B. Sanyal, J. Physics-Condensed Matter **24**, 433504 (2012).
[14] N. A. Deskins, R. Rousseau, and M. Dupuis, J. Phys. Chem. C **115**, 7562 (2011).
[15] B. J. Morgan and G. W. Watson, Surf. Sci. **601**, 5034 (2007).
[16] C. M. Yim, C. L. Pang, and G. Thornton, Phys. Rev. Lett. **104**, 036806 (2010).
[17] P. Krüger, S. Bourgeois, B. Domenichini, H. Magnan, D. Chandesris, P. Le Fèvre, A. Flank, J. Jupille, L. Floreano, A. Cossaro, A. Verdini, and A. Morgante, Phys. Rev. Lett. **100**, 055501 (2008).
[18] Y. B. He, O. Dulub, H. Z. Cheng, A. Selloni, and U. Diebold, Phys. Rev. Lett. **102**, (2009).
[19] M. Setvín, U. Aschauer, P. Schreiber, Ye-Fei Li, W. Hou, M. Schmied, A. Selloni, and U. Diebold, Science, **341**, 988 (2013).
[20] V. Henrich and R. Kurtz, Phys. Rev. B **23**, 6280 (1981).
[21] R. Sanjinés, H. Tang, H. Berger, F. Gozzo, G. Margaritondo, and F. Lévy, J. Appl. Phys. **75**, 2945 (1994).
[22] A. G. Thomas, W. R. Flavell, A. K. Mallick, A. R. Kumarasinghe, D. Tsoutsou, N. Khan, C. Chatwin, S. Rayner, G. C. Smith, R. L. Stockbauer, S. Warren, T. K. Johal, S. Patel, D. Holland, A. Taleb, and F. Wiame, Phys. Rev. B **75**, (2007).
[23] S. Wendt, P. T. Sprunger, E. Lira, G. K. H. Madsen, Z. S. Li, J. O. Hansen, J. Matthiesen, A. Blekinge-Rasmussen, E. Laegsgaard, B. Hammer, and F. Besenbacher, Science **320**, 1755 (2008).
[24] C. L. Pang, R. Lindsay, and G. Thornton, Chem. Soc. Rev. **37**, 2328 (2008).
[25] C. Sánchez-Sánchez, M. G. Garnier, P. Aebi, M. Blanco-Rey, P. L. de Andres, J. A. Martín-Gago, and M. F. López, Surf. Sci. **608**, 92 (2013).
[26] A. G. Thomas, W. R. Flavell, A. R. Kumarasinghe, A. K. Mallick, D. Tsoutsou, G. C. Smith, R. Stockbauer, S. Patel, M. Gratzel, and R. Hengerer, Phys. Rev. B **67**, 3 (2003).
[27] M. A. Henderson, W. S. Epling, C. H. F. Peden, and C. L. Perkins, J. Phys. Chem. B **107**, 534 (2003).
[28] N. Fairley, CasaXPS Software
[29] K. Onda, B. Li, and H. Petek, Phys. Rev. B **70**, 045415 (2004).
[30] The very slight change in intensity is believed to be an artefact of the peak fitting.
[31] M. A. Henderson, Surf. Sci. Rep. **66**, 185 (2011).

[32] M. Setvin, X. Hao, B. Daniel, J. Pavelec, Z. Novotny, G. S. Parkinson, M. Schmid, G. Kresse, C. Franchini, and U. Diebold, Angew. Chemie Int. Ed. **53**, 4714 (2014).

[33] L. Forro, O. Chauvet, D. Emin, L. Zuppiroli, H. Berger, and F. Lévy, J. Appl. Phys. **75**, 633 (1994).

# Supplementary Information

***S.1.****: The valence band of Rutile TiO$_2$ (110) and Anatase TiO$_2$ (101) recorded at 47 eV. The valence band structure arises from O 2p states with some Ti 3d and 4sp character.*

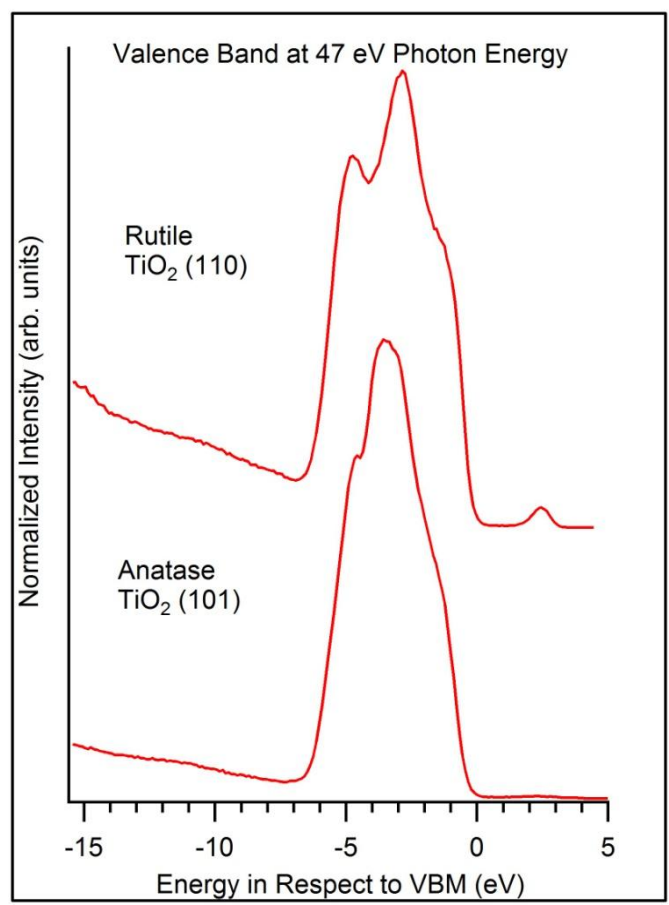

***S.2.:*** *Determination of Rutile (110) TiO$_2$ peak ratio and position in relation to the valence band maximum (VBM).*

| | Ratio of Peaks | | | Position of Peaks in Relation to VBM | |
|---|---|---|---|---|---|
| | **Area** | **Area** | | **eV w.r.t. VBM** | **eV w.r.t. VBM** |
| Rep 1 | 3730.9 | 472.8 | Rep 1 | -2.11 | -1.47 |
| Rep 2 | 3880.9 | 363.5 | Rep 2 | -2.09 | -1.34 |
| Rep 3 | 3774.6 | 457.1 | Rep 3 | -2.12 | -1.47 |
| Rep 4 | 3802.6 | 451.7 | Rep 4 | -2.07 | -1.42 |
| Rep 5 | 3935.5 | 363.3 | Rep 5 | -2.12 | -1.39 |
| Rep 6 | 3952.0 | 315.6 | Rep 6 | -2.07 | -1.17 |
| Rep 7 | 3581.5 | 456.0 | Rep 7 | -2.09 | -1.41 |
| Rep 8 | 3956.5 | 348.6 | Rep 8 | -2.11 | -1.27 |
| **Mean** | 3826.8 | 403.6 | **Mean** | **-2.1** | **-1.4** |
| **S.D.** | 130.9 | 61.8 | **S.D.** | 0.0 | 0.1 |
| **R.S.D.** | 3.4 | 15.3 | **R.S.D.** | -1.0 | -7.6 |
| **Ratio** | **90.5** | **9.5** | | | |

To determine the energy calibration and ratio of the R-TiO$_2$ (110) defect peaks, eight spectra were taken and the mean energy with respect to the VBM was calculated. The position in relation to the VBM was +1.4 eV (binding energy: 1.7 eV) and +2.1 eV (binding energy: 1.0 eV) in a ratio of 10:90 with RSD % of the peak areas of 15.3 % and 3.4 % respectively, the former being significantly larger due to the small area of the peak. The low RSDs of the peak areas and no obvious trends in the data means the states are stable and there is no notable damage to the surface caused by the beam.

***S.3.:*** *Ti 3p peak for Rutile TiO$_2$ (110) [a]. A shirley background was used and the peaks were fitted with Voigt curves (70% Gaussian: 30% Lorentzian) using CASA XPS software [b].*

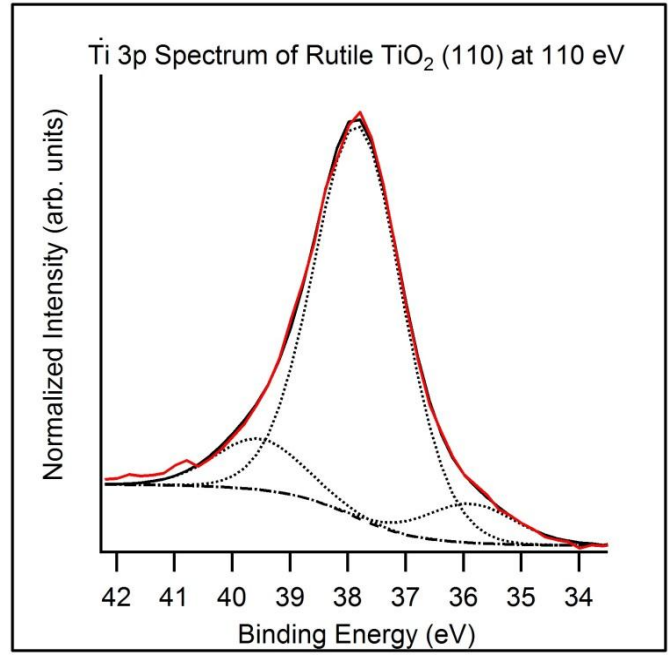

| Peak Assignment | Binding Energy eV | % |
|---|---|---|
| Bulk | 37.8 | 80.0 |
| Ti$^{3+}$ | 35.7 | 6.9 |
| Surface hydroxyls | 40.2 | 13.1 |

***S.4.:*** *Determination of Anatase TiO$_2$ (101) position in relation to the valence band maximum (VBM).*

|  | Energy in relation to VBM (eV) | | |
|---|---|---|---|
|  | **Peak i** | **Peak ii** | **Peak iii** |
| Rep 1 | -2.18 | -1.5 | -0.75 |
| Rep 2 | -2.23 | -1.54 | -0.89 |
| Rep 3 | -2.27 | -1.55 | -0.81 |
| Rep 4 | -2.26 | -1.53 | -0.78 |
| Rep 5 | -2.29 | -1.57 | -0.81 |
| Rep 6 | -2.27 | -1.57 | -0.86 |
| Rep 7 | -2.3 | -1.57 | -0.85 |
| Rep 8 | -2.33 | -1.63 | -0.85 |
| Rep 9 | -2.18 | -1.5 | -0.75 |
| Rep 10 | -2.23 | -1.54 | -0.89 |
| Rep 11 | -2.27 | -1.55 | -0.81 |
| Rep 12 | -2.26 | -1.53 | -0.78 |
| Rep 13 | -2.29 | -1.57 | -0.81 |
| Rep 14 | -2.27 | -1.57 | -0.86 |
| Rep 15 | -2.3 | -1.57 | -0.85 |
| Rep 16 | -2.33 | -1.63 | -0.85 |
| **Mean** | **-2.27** | **-1.56** | **-0.83** |
| **SD** | **0.05** | **0.04** | **0.05** |
| **RSD %** | **-2.01** | **-2.44** | **-5.57** |

To determine the energy calibration of the A-TiO$_2$ (101) defect peaks, sixteen spectra were taken and the mean energy with respect to the VBM was calculated. The position in relation to the VBM was +0.8 eV (binding energy: 2.3 eV), +1.6 eV (binding energy: 1.0 eV) and +2.3 eV (binding energy: 1.7 eV). The low RSDs of the energy positions indicates consistent fitting.

**REFERENCES FOR SUPPLEMENTARY INFO**

[a] A. G. Thomas, W. R. Flavell, A. K. Mallick, A. R. Kumarasinghe, D. Tsoutsou, N. Khan, C. Chatwin, S. Rayner, G. C. Smith, R. L. Stockbauer, S. Warren, T. K. Johal, S. Patel, D. Holland, A. Taleb, and F. Wiame, Phys. Rev. B **75**, (2007).

[b] Fairley, N. CasaXPS: V2.3.16.